\documentstyle[twoside,fleqn,espcrc2]{article}

\newcommand{\AmS}{{\protect\the\textfont2
  A\kern-.1667em\lower.5ex\hbox{M}\kern-.125emS}}
 
\title{Cosmological Quantum String Vacua and String-String Duality }

\author{SOO-JONG REY\address{Physics Department, Seoul National University,
Seoul 151-742 KOREA}%
        \thanks{
Research supported in part by U.S.NSF-KOSEF Bilateral Grant, KOSEF
Purpose-Oriented Research Grant 94-1400-04-01-3 and SRC-Program,
KRF Nondirected Research Grant 81500-1341 and International Collaboration
Grant, and Ministry of Education BSRI Grant 94-2418.} }

\begin{document}
\setlength{\arraycolsep}{0pt}
\begin{titlepage}
\begin{flushright} SNUTP 96-076\\ {\tt hep-th/9607148}
\end{flushright}
\vfill
\begin{center}
{\large\bf Cosmological Quantum String Vacua and String-String 
Duality${}^\dagger$}\\
\vskip 7.mm
{\large \sl Soo-Jong Rey}\\
\vskip 0.5cm                                                      
{\em Physics Department \& Center for Theoretical Physics} \\
{\em Seoul NationaL University}\\
{\em Seoul 151-742 KOREA} 
\end{center}
\vfill
\begin{center}
{\bf ABSTRACT}
\end{center}
\begin{quote}
Implications of string-string dualities to cosmological string vacua
are discussed. Cosmological vacua of classical string theories comprise of
disjoint classses mapped one another by scale-factor T-duality. Each classes
are, however, afflicted with initial/final cosmological singularities.
It is argued that quantum string theories and string-string dualities
dramatically resolve these cosmological singularities out so that disjoint
classical cosmological vacua are continuously connected in a unified manner.
A natural inflationary cosmology follows from this.

\vfill      \hrule width 5.cm
\vskip 2.mm
{\small\small
\noindent $^\dagger$ Invited talk given at the Fourth International 
Conference on Supersymmetry (SUSY '96), College Park, May 26 - June 1, 1996; 
To be published in the proceedings.}  
\end{quote}
\end{titlepage}

\begin{abstract}
Implications of string-string dualities to cosmological string vacua 
are discussed. Cosmological vacua of classical string theories comprise of 
disjoint classses mapped one another by scale-factor T-duality. Each classes
are, however, afflicted with initial/final cosmological singularities.
It is argued that quantum string theories and string-string dualities 
dramatically resolve these cosmological singularities out so that disjoint 
classical cosmological vacua are continuously connected in a unified manner.
A natural inflationary cosmology follows from this.
\end{abstract}

\maketitle
\setcounter{footnote}{0}

Tremendous progress in recent years to nonperturbative string 
theory~\cite{duality} has now shown compelling evidences that all known 
perturbative string theories (type-I, heterotic, type II) are equivalent 
nonperturbatively up to duality transformations~\cite{stringduality}. 
String dualities have been studied so far, 
however, only for compactifications to Minkowski spacetime. 
Cosmological string vacua are described by an `adiabatic motion' of spacetime 
geometry in the string moduli space, viz. a fibration of three-geometry, 
string coupling and compactified space over `cosmological time'. 
It is then of immediate interest what the string dualities might tell us for 
string cosmology. In particular we would like to know if the string dualities 
can offer a resolution to cosmological singularity in a similar spirit 
to the conifold transition~\cite{andy}.  
In this talk, based on a simple two-dimensional compactification, 
I exemplify that string dualities offer an interesting cosmological quantum
string vacua with an inflation but no initial or final 
singularities~\cite{mine}.

String theories exhibit generically two interesting classical cosmological 
vacua~\cite{vene}: 
(1) Friedmann-Robertson-Walker (FRW)-like vacua of decelerating expansion 
$\ddot a(t) \!\! < \!\! 0,\, \dot a(t)\! \! > \! \! 0$, 
(2) super-inflationary vacua of accelerating expansion 
$\ddot a(t) \!\! > \!\! 0, \, \dot a(t) \!\! > \!\! 0$, related each 
other under scale factor T-duality accompanied by a time-reversal. The two
vacua are, however, afflicted with initial/final singularities of spacetime
geometry. In fact, scale factor T-duality (accompanied by a time-reversal)
relates the two vacua, but offers no resolution of the singularities.
Because of this, the classical string cosmology is afflicted by a 
`graceful exit problem'~\cite{gracefulexit}.

Nature of the problem and quantum resolution thereafter can be
treated exactly in two-dimensional sting compactifications. A given 
compactification is specified by spacetime geometry, dilaton and various 
compactification moduli, and is described by the same action as the dilaton 
gravity~\cite{cghs} with vanishing central charge deficit 
(cosmological constant)
\begin{equation}
e^{-1} L_0 = [e^{-2 \phi} (R + 4 (\nabla \phi)^2)
-{1 \over 2} (\nabla {\vec f})^2 ]
\label{2daction}
\end{equation}
where $\phi$ and ${\vec f}$ refer to dilaton and $N$-component 
Ramond-Ramond scalar field. In conformal gauge
$ds^2 = - e^{2 \rho} d x_+ d x_-; \hskip0.2cm x_\pm = t \pm x, 
\hskip0.2cm \partial_\pm = {1 \over 2} (\cdot \pm \prime)$, 
the Lagrangian is expressed in terms of field variables
$\Phi = e^{-2 \phi}, \,\, \Sigma = 2 \kappa (\phi - \rho)$ as 
\begin{equation}
L_0 = {1 \over 2 \kappa} \big( \partial_+ \Phi \partial_- \Sigma
+\partial_- \Phi \partial_+ \Sigma \big) + 
{1 \over 2} \partial_+ {\vec f} \cdot \partial_- {\vec f}
\label{newaction}
\end{equation}
supplemented with constraint equations
\begin{equation}
T_{\pm\pm} = {1 \over 2} (\partial_\pm {\vec f})^2 
+ \partial_\pm^2 \Phi + {1 \over \kappa} \partial_\pm \Phi 
\partial_\pm \Sigma = 0.
\label{newconstr}
\end{equation}
A parameter $\kappa$ is introduced to keep track of perturbation expansions.

In the classical limit $\kappa \rightarrow 0$, the theory
exhibit two useful symmetries $\Phi \rightarrow \Phi +  \epsilon_\Sigma, 
\,\, \Sigma \rightarrow \Sigma$ and $\Sigma \rightarrow \Sigma + \epsilon_\Phi,
\Phi \rightarrow \Phi$, thus exactly soluble.
The most general cosmological solution to equations of motion 
$\ddot \Phi = \ddot \Sigma = 0, \,\, {\vec f} = {\rm constant}$
are
\begin{equation}
-{1 \over 2 \kappa} \Sigma  = Q_\Sigma t + A, \,\,\, \Phi = Q_\Phi t + B
\end{equation}
subject to Eq.(\ref{newconstr}) $T_{\pm\pm} = 2 Q_\Phi Q_\Sigma = 0. $

For the vacua with $Q_\Phi \!\! \ne \!\! 0: \rho \!\! = \!\! 
\phi + \log 2 {\tilde M}, \, e^{-2\phi} =  - 8 {\tilde M} t$, 
the spacetime is given by
\begin{equation} 
(ds)^2 = - [d \tau^2 - ({{\tilde M} \over - \tau})^2 dx^2]; 
\hskip0.3cm -\infty < \tau \le 0
\end{equation}
with cosmic time $\tau := - (- 2 {\tilde M} t)^{1/2}$.
The comoving scale factor $a(\tau) \propto 1/(-\tau)$ shows 
that this vacua describes a super-inflationary evolution. Dilaton
grows large as $\phi = - \log (-2\tau)$.

For the other vacua with $Q_\Sigma \ne 0: \,\, \rho = \phi + M t , \,\, 
e^{-2\phi} = M^{-2}$, the spacetime is given by
\begin{equation} 
(ds)^2 = - [ d\tau^2 - (M \tau)^2 dx^2]; 
\hskip0.3cm 0 \le \tau < \infty 
\end{equation}
with $\tau: = \exp Mt, \, 0  \le \tau < \infty$.
This linearly expanding unverse with a constant dilaton is a 
two-dimensional counterpart of radiation-dominated FRW-type vacua. 

Both vacua are afflicted by divergent curvature $R = (2/|\tau|)^2 
\rightarrow \infty$ at $\tau = 0$, viz., cosmological initial or final 
singularities. While the two vacua are related each other by scale factor 
$T$-duality and time-reversal: $-{\tilde M} / \tau \rightarrow M \tau$,
patching them together using the T-duality does not remove the singularity 
since both curvature and string coupling remain singular and discontinuous 
across the junction $\tau = 0$. 
This difficulty persists even after higher-derivate $\alpha'$ 
corrections are included (also in $D=4$); it is the the aformentioned 
\sl graceful exit problem \rm~\cite{gracefulexit}.

There is a good reason, however, to expect that the situation might be changed 
for \sl quantum \rm string vacua. Near the T-duality junction, 
$\tau \approx 0$, the string coupling diverges, 
hence, back reaction will become important~\cite{anto}.
At one-loop, quantum correction is provided by conformal anomaly of 
$N$-component matter, reparametrization ghosts, dilaton and conformal modes. 
For $N>5$ supersymmetric compactifications (corresponding to $D=4, N> 2$ ones),
the quantum correction is saturated at one-loop, hence, the above consideration
yields an exact quantum theory.
After adding a local counterterm to retain the classical symmetries for 
exact solvability~\cite{bilalcallan}, the quantum effective Lagrangian 
is given by
\begin{equation}
L_{\rm eff} = L_0 - {\kappa \over 2} \big[ R {1 \over \Box } R + 2 \phi R \big]
\label{effaction}
\end{equation}
where $\kappa = (N-24)/24$. 
In terms of quantum-corrected fields 
$\Sigma \!\!=\!\! 2 \kappa (\phi - \rho); \,\,\, \Phi \!\!= \!\! 
e^{-2 \phi }  + \kappa \rho$
$L_{|rm eff}$ is exactly the same as classical $L_o$, 
but the constraint equation is corrected to
\begin{equation}
T_{\pm\pm} - \kappa \big({1 \over 2} \partial_\pm \Sigma + t_\pm (x^\pm) \big)
= 0.
\label{quantumconstr}
\end{equation}
The first integrals $t_\pm(x^\pm)$ coming from nonlocality of conformal anomaly
are fixed $t_\pm = 0$ to yield the Minkowski vacuum for a constant dilaton.

It turns out that physically acceptable cosmological \sl quantum \rm vacua 
result in only for negative conformal anomaly  $\kappa < 0$, $N < 24$:
\begin{equation}
{1 \over 2 |\kappa|} \Sigma = Q_\Sigma t + A ; \,\,\, \Phi = Q_\Phi t + B
\end{equation}
subject to  Eq.(\ref{quantumconstr}) $ 0 = \kappa t_\pm = Q_\Phi Q_\Sigma$.

The first quantum vacua with $Q_\Phi \ne 0$ that exhibited classically 
a super-inflation is given by
$\rho = \phi + \log 2 {\tilde M}  ; \,\, 
e^{-2 \phi} - |\kappa| \rho = - 8 {\tilde M} t. $
Most importantly quantum correction now extends the cosmic evolution beyond 
the classical range $-\infty < t \le 0$ into $-\infty < t < + \infty$ so long 
as $\kappa < 0$. Noting that $\phi$ is a monotonic function of $t$ the string 
coupling $g_{\rm st} = e^\phi$ may be viewed as a built-in clock.  As expected,
near $ t \approx - \infty$, the vacua exhibit a classical, super-inflation
$\dot a(\tau) \!\!>\!\! 0, \ddot a(\tau) \!\!>\!\! 0$.
On the other hand, as $t \rightarrow + \infty$, 
$\dot a(\tau) > 0$ but $\ddot a(\tau) < 0$ and approaches asymptotically to
\begin{equation}
(ds)^2 \rightarrow 
- [d \tau^2 - ({8 {\tilde M} \over |\kappa|} \tau)^2 dx^2];
\hskip0.3cm \phi \rightarrow  \log \tau,
\end{equation}
where $\tau \approx (|\kappa| / 8 {\tilde M}) 
\exp (8 {\tilde M} t / |\kappa|)$.
The initially super-inflated universe has now decelerated into FRW-like 
expansion with a linear dilaton. More insight into the interpolating behavior 
of the quantum vacua can be gained from the scalar curvature
as a function of string coupling $g_{\rm st}$:
$R = 16 g_{\rm st}^2 / (1 + |\kappa| g_{\rm st}^2/2)^3.$
The curvature vanishes at past/future infinity $g_{\rm st} 
\rightarrow 0/\infty$ but reaches a finite maximum at T-duality junction 
$ \tau = 0$. Hence, the classical singularity at $\tau = 0$ is now completely 
erased out and the inflation has ended gracefully!

Interestingly, the second quantum vacua with $Q_\Sigma \ne 0: 
\rho = \phi + Mt ; \,\,\, e^{-2 \phi} - |\kappa|  \rho = M^{-2}$
that corresponded classically to a distinct, FRW-like vacua
is now related to the first quantum vacua by S-duality.
At $ t \rightarrow + \infty$, the vacua exhibit expected classical FRW-like  
evolution
\begin{equation}
(ds)^2 \rightarrow  -[d\tau^2 - ({\tau \over |\kappa|})^2 dx^2]; 
\phi \rightarrow {\rm const}.
\end{equation}
At $ t \rightarrow - \infty$, however, the vacua 
\begin{equation}
(ds)^2 \rightarrow 
 -[d\tau^2 - \exp(2 e^{2M\tau}) dx^2]; \phi \rightarrow  - M \tau,
\end{equation}
evolves initially as a Minkowski spacetime with a linear dilaton 
then into  super-inflation $\dot a(\tau) > 0$, $\ddot a(\tau) > 0$,
$1/H \approx \exp(-2M\tau) \rightarrow 0$.
Scalar curvature expressed as a funtion of $g_{\rm st}$ takes a 
suggestive nonperturbative form associated with particle production 
$R \propto \exp (-{2 \over |\kappa|} {1 \over g_{\rm st}^2})
g_{\rm st}^4 / ( 1 + |\kappa| g_{\rm st}^2 / 2)^3$. It vanishes at 
asymptotic past/future infinities $g_{\rm st} \rightarrow \infty/0$ but 
approaches a finite positive maximum at T-duality junction $\tau = 0$.
We thus conclude that the second quantum vacua is S-dual to the first one 
and both resolve classical singularities and exit super-inflation 
epoch gracefully. We again emphasize that, no matter how small $g_{\rm st}$ 
and $|\kappa|$ were, the quantum cosmological vacua behave dramatically 
different from the classical cosmological vacua, similar to ~\cite{andy}.

One common aspect of both quantum vacua is that the string coupling
$g_{\rm st} = e^\phi$ evolves monotonically: either from weak to strong for 
the first vacua or from strong to weak coupling for the second. 
The two \sl quantum \rm cosmological vacua are S-dual each other and
interpolate between the two classical cosmological vacua related by 
scale-factor $T$-duality and time-reversal.  
This is in accord with the emerging picture of string-string dualities:
string coupling evolution from weak to strong or vice versa 
is equivalent to an interpolation between two weakly coupled phases of 
dual string pairs~\cite{stringduality} such as type-I and heterotic and 
type-II/M-theory and heterotic strings. 
Interestingly the scale-factor T-duality junction $\tau = 0$ is precisely 
where $g_{\rm st}$ becomes order unity, hence, where worldsheet and string 
loop perturbation expansions in both dual pair string threories break down. 
Evidently this is where quantum back reaction is most pronounced and the
classical singularities are erased completely.

Preliminary study indicates that higher dimensional compactifications also 
exhibit similar behavior: singularity-free, cosmological quantum string vacua
all connected and unified via S- and T-duality transformations. 
It is undoubtedly gratifying that string dualities seem to offer solutions 
to the outstanding problems of classical string cosmology.


\begin{thebibliography} {9}

\bibitem{duality}
A. Font, L. Ibanez, D. Lust and F. Quevedo, Phys. Lett. \bf 249B,
\rm 35 (1990);
S.-J. Rey, Phys. Rev. \bf D43, \rm 526 (1991);
M.J. Duff and J.X. Lu, Nucl. Phys. \bf B357, \rm 534 (1994);
A. Sen and J.H. Schwarz, Phys. Lett. \bf 312B, \rm 105 (1993);
C.M. Hull and P.K. Townsend, Nucl. Phys. \bf B438, \rm 109 (1995);
E. Witten, Nucl. Phys. \bf B443, \rm 85 (1995).

\bibitem{stringduality}
J. Polchinski and E. Witten, Nucl. Phys. \bf B460, \rm 525 (1996);
Horava and E. Witten, Nucl. Phys. \bf B460, \rm 506 (1996); \sl
ibid. \rm hep-th/9603142 (unpublished).
 
\bibitem{andy} A. Strominger, Nucl. Phys. \bf B451, \rm 96 (1995);
B.R. Greene, D.R. Morrison and A. Strominger, Nucl. Phys. \bf B451, \rm
109 (1995).

\bibitem{mine} S.-J. Rey, \tt hep-th/9605176 \rm (1996).

\bibitem{vene} G. Veneziano, Phys. Lett. \bf 265B, \rm 287 (1991);
M. Gasperini and G. Veneziano, Astrop. Phys. \bf 1, \rm 317 (1993).

\bibitem{gracefulexit} R. Brustein and G. Veneziano, Phys. Lett.
\bf 329B, \rm 429 (1994); N. Kaloper, R. Madden and K.A. Olive,
Nucl. Phys. \bf B452, \rm 677 (1995); R. Easther, K. Maeda and D. 
Wands, Phys. Rev. \bf D53, \rm 4247 (1996).

%\bibitem{universal} M. Gasperini and G. Veneziano, Mod. Phys. Lett. \bf A8, \rm 3701 (1993); Phys. Rev. \bf D50, \rm 2519 (1994).

\bibitem{cghs} C.G. Callan, S. Giddings, J. Harvey and A. Strominger,
Phys. Rev. \bf D45, \rm 1005 (1992).

\bibitem{anto} I. Antoniadis, J. Rizos and K. Tamvakis, Nucl. Phys.
\bf B415, \rm 497 (1994); E. Kiritsis and C. Kounnas, Phys. Lett.
\bf B331, \rm 51 (1994).

\bibitem{bilalcallan} A. Bilal and C.G. Callan, Nucl. Phys. \bf B394, \rm 
73 (1993); J.G. Russo, L. Susskind and L. Thorlacius, Phys. Rev. \bf
D46, \rm 3444 (1992); Phys. Rev. \bf D47, \rm 533 (1993).

\end{thebibliography}
\end{document}